\documentclass[12pt]{article}
\usepackage{epsf,amsfonts,amssymb,epsfig,amsmath}
\usepackage{color}

\renewcommand{\baselinestretch}{1.2}

\begin{document}

\makeatletter \@addtoreset{equation}{section} \makeatother
\renewcommand{\theequation}{\thesection.\arabic{equation}}
\renewcommand{\thefootnote}{\alph{footnote}}

\begin{titlepage}

\begin{center}
\hfill {\tt SNUTP10-002}\\
\hfill {\tt arXiv:1001.3153}

\vspace{2.5cm}

{\large\bf Supersymmetric vacua of mass-deformed M2-brane theory}

\vspace{2.5cm}

\renewcommand{\thefootnote}{\alph{footnote}}

{\large Hee-Cheol Kim and Seok Kim}

\vspace{1cm}

\textit{Department of Physics and Astronomy \& Center for
Theoretical Physics,\\
Seoul National University, Seoul 151-747, Korea.}\\

\vspace{0.7cm}

E-mails: {\tt heecheol1@gmail.com, skim@phya.snu.ac.kr}

\end{center}

\vspace{2.5cm}

\begin{abstract}

We count the supersymmetric vacua of mass-deformed $\mathcal{N}\!=\!6$ $U(N)\!\times\!U(N)$ Chern-Simons-matter theory by calculating the Witten index. When the Chern-Simons level $k$ is $1$, our result perfectly agrees with that from the gravity dual, given by partitions of $N$. For general $k$, our index generalizes partitions of $N$, including additional degrees. We also comment on non-relativistic superconformal theories constructed from this model.

\end{abstract}

\end{titlepage}

\renewcommand{\thefootnote}{\arabic{footnote}}

\setcounter{footnote}{0}

\renewcommand{\baselinestretch}{1}

\tableofcontents

\renewcommand{\baselinestretch}{1.2}

\section{Introduction and discussions}

There has been a lot of recent progress in microscopic understanding of M2-branes from  Chern-Simons-matter theories. See \cite{Bagger:2006sk,Aharony:2008ug} and an extensive list of references thereof. Much of the work was on the conformal field theories and their AdS$_4$ gravity duals. It is also interesting to study nonconformal quantum field theories in 3 dimensions for M2-branes. A simple nonconformal model which has drawn some attention is the $U(N)_k\!\times\!U(N)_{-k}$ $\mathcal{N}\!=\!6$ Chern-Simons-matter theory with mass-deformation \cite{Hosomichi:2008jb,Gomis:2008vc}. From the M-theory perspective, this is the worldvolume theory of $N$ parallel M2-branes in $\mathbb{R}^8/\mathbb{Z}_k$ with a nonzero 4-form field on two $\mathbb{R}^4$ factors of $\mathbb{R}^8$. This theory compactified on a 2-torus also describes the discrete light-cone quantization (DLCQ) of type IIB strings in the maximally supersymmetric plane wave \cite{Sethi:1997sw}.

The mass-deformed theory has many discrete vacua. The classical vacua preserving supersymmetry were studied in \cite{Gomis:2008vc}. They are interpreted as M5-branes stretched along $\mathbb{R}^{2\!+\!1}$ and polarized into 3-spheres in one of the two $\mathbb{R}^4$ parts in $\mathbb{R}^8$ \cite{Bena:2000zb}. With 2-torus compactification, these vacua are dualized to polarized half-BPS D3-branes in the plane wave \cite{Bena:2004qv}.

For $k\!=\!1$, M2-branes in $\mathbb{R}^8$ show enhanced $\mathcal{N}\!=\!8$ supersymmetry. In this case, the supersymmetric vacua have been extensively studied from the gravity side, relying on the constraints from $16$ supersymmetries \cite{Lin:2004nb,Bena:2004jw}. These half-BPS geometries are governed by fermion droplets on a cylinder, with a boundary condition that one end of the cylinder is filled. By a semi-classical quantization, one obtains the partition function of supersymmetric vacua. The result $I_{1N}$ for $k\!=\!1$ is given by partitions of $N$, whose generating function is
\begin{equation}\label{partition}
  I_1(q)\equiv\sum_{N=1}^\infty I_{1N}q^N=\prod_{n=1}^\infty\frac{1}{1-q^n}\ .
\end{equation}
The same quantity can be obtained from the gauge theory dual of type IIB strings in the DLCQ plane wave \cite{Mukhi:2002ck}. The result (\ref{partition}) is expected to be correct for both uncompactified M2-branes and those on $T^2$.

There are various ways of understanding (\ref{partition}). In particular, from the gravity side, this can be understood as coming from quantizing polarized D3-brane giant gravitons in the type IIB plane wave background after $T^2$ compactification. From the classical probe brane perspective, branes expanded in two $\mathbb{R}^4$ yield two classes of solutions: one class called `giant graviton' and another `dual giant graviton' \cite{McGreevy:2000cw}. The prescription of getting the correct result (\ref{partition}), sometimes called `giant graviton complementarity,' states that the true quantum degeneracy of half-BPS states (\ref{partition}) is given by (naively) quantizing either giant gravitons or dual giant gravitons, but not both. This notion is generalized to various $\frac{1}{8}$-BPS sectors \cite{Biswas:2006tj,Grant:2008sk}.

Curiously, the classical vacua in the mass-deformed $\mathcal{N}\!=\!6$ Chern-Simons theory are similar to the classical polarized brane solutions, in that there are two possible $\mathbb{R}^4$ in $\mathbb{R}^8$ in which the scalar fields can be turned on \cite{Gomis:2008vc}. See section 2. The degeneracy of quantum vacua should be compatible with the result obtained from the complementarity prescription, which now has to be derived microscopically from quantum field theory.

In this paper, we count the quantum supersymmetric vacua of this theory for general $k$. Since the strongly interacting regime, say $k\!=\!1$, is included, we are naturally led to study the Witten index for these vacua, counting the number of bosonic vacua minus that of fermionic ones. At least for $k\!=\!1$, the gravity states are all bosonic so that the partition function (\ref{partition}) should be equal to the index.

To compute the index, it turns out to be possible (and helpful) to deform the Chern-Simons-matter theory by turning on Yang-Mills kinetic terms for the gauge fields. We introduce such a deformation preserving an $\mathcal{N}\!=\!2$ part of the full supersymmetry. Following \cite{Witten:1999ds}, we also compactify this theory on a small 2-torus to reduce our study to a mechanical index. A crucial part of our analysis will come from the quantum dynamics of $\mathcal{N}\!=\!2$ Yang-Mills Chern-Simons theory with the gauge group unbroken by a classical supersymmetric vacuum. Indices for such theories are computed in \cite{Ohta:1999iv,Acharya:2001dz}, following \cite{Witten:1999ds}.\footnote{Deforming to $\mathcal{N}\!=\!3$ Yang-Mills theory might be physically more natural, considering the origin of M2-brane theories from $\mathcal{N}\!=\!3$ D-brane systems. The resulting index would be the same: technically, this is because $\mathcal{N}\!=\!2$ and $3$ Yang-Mills Chern-Simons theories have same index \cite{Ohta:1999iv}.} Strictly speaking, we are considering the vacua on $T^2$, or half-BPS states in the DLCQ plane wave. However, from various indirect field theory arguments in \cite{Witten:1999ds} as well as successful applications of Yang-Mills Chern-Simons index on $T^2$ to D-branes living on uncompactified $\mathbb{R}^{2\!+\!1}$ \cite{Ohta:1999iv,Acharya:2001dz}, we believe that our calculation also provides the vacuum counting for M2-branes living on $\mathbb{R}^{2\!+\!1}$.

Our main result is the index $I_{kN}$ of supersymmetric vacua for general $k$ ($\neq 0$):
\begin{equation}\label{general-partition}
  I_k(q)\equiv\sum_{N=0}^\infty I_{kN}q^N=
  \prod_{n=1}^\infty\frac{1}{(1-q^n)^{|k|}}\ \sum_{\substack{n_1,n_2,\cdots,n_{|k|}=
  -\infty\\(n_1+n_2+\cdots+n_{|k|}=0)}}^\infty q^{\frac{1}{2}\left(n_1^2+n_2^2+\cdots+n_{|k|}^2\right)}\ .
\end{equation}
For $k=1$, the second factor of restricted infinite sum reduces to $1$, so that the gravity result (\ref{partition}) is immediately reproduced. See section 3 for its derivation. An important reason enabling this perfect agreement, reducing apparent richness of classical vacua, is the dynamical supersymmetry breaking of Yang-Mills Chern-Simons theory \cite{Witten:1999ds} when the unbroken gauge group is large compared to the Chern-Simons level $k$. See section 3 for the details.

Our results for $k\geq 2$ do not seem to be well understood in the gravity dual. In any case, since we started from $\mathbb{R}^8/\mathbb{Z}_k$, the resulting geometry cannot be completely smooth. For instance, we find the following expressions for some low values of $k$,
\begin{eqnarray}\label{higher-partition}
  I_2(q)&=&\prod_{n=1}^\infty\frac{(1+q^n)(1+q^{2n\!-\!1})^2}{1-q^n}\\
  I_3(q)&=&\prod_{n=1}^\infty\frac{(1\!+\!q^n)^2(1\!+\!q^{2n}\!+\!q^{4n})}{1-q^n}\left[
  \prod_{n=1}^\infty(1\!+\!q^{2n\!-\!1})(1\!+\!q^{6n\!-\!3})^2+
  4q\prod_{n=1}^\infty(1\!+\!q^{2n})(1\!+\!q^{6n})^2\right]\nonumber
\end{eqnarray}
after massaging the infinite sum in (\ref{higher-partition}). There is always a factor like (\ref{partition}). All other infinite products appearing in the numerator allude to contribution from additional degrees.
It would be very interesting to see if these can be understood from the gravity dual. For example, since the case with $k\!=\!2$ also preserves $16$ supersymmetry, the local structure of the gravity solutions should be the same as \cite{Lin:2004nb}. It seems to us that a factor of $S^3\times S^3$ appearing in the half-BPS solution should be modded out by a $\mathbb{Z}_2$ in the following way. Near a boundary of black and white regions in the droplet, one locally encounters the appearance of $\mathbb{R}^8$ for $k\!=\!1$, by combining $S^3\times S^3$ with two radial directions. The $\mathbb{Z}_2$ we mentioned above should replace this by $\mathbb{R}^8/\mathbb{Z}_2$ for $k\!=\!2$. Even for general $k$ with $\mathcal{N}\!=\!6$ supersymmetry, it might be that the relevant gravity solutions come with $\mathbb{Z}_k$ modding of the half-BPS solutions. The common appearance of the factor (\ref{partition}) in (\ref{higher-partition}) seems to imply that the cases with $k\!\geq\!2$ share some common features with the case with $k\!=\!1$.

For $k\!=\!1$, we also provide a concrete 1 to 1 map between the supersymmetric vacua and the gravity solutions in \cite{Lin:2004nb}. Based on this map, one would be able to pick a supersymmetric vacua whose gravity dual has small curvature and consider interesting holographic calculations. We leave such studies to the future.

A motivation for previous studies of mass-deformed Chern-Simons-matter theories was to find a microscopic realization of non-relativistic holography \cite{Nakayama:2009cz,Nakayama:2009ed} with the so-called Schr\"{o}dinger symmetry. In \cite{Nakayama:2009cz}, non-relativistic superconformal theories were constructed starting from the classical vacuum in which no scalar fields are turned on (i.e. no polarization). See also \cite{Kwon:2010ev}. In our study, this vacuum dynamically breaks supersymmetry unless $N\leq k$. This could be signaling that the holography with super-Schr\"{o}dinger symmetry obtained from this model would be highly `stringy' from the gravity side, in that the 't Hooft coupling $\frac{N}{k}\leq 1$ is small. See also \cite{Nakayama:2009ed} for similar discussions.

There have also been studies of supersymmetric gravity solutions with Schr\"{o}dinger symmetry. It was observed that solutions with $14$ or more supersymmetries, needed for the non-relativistic superconformal theory obtained from mass-deformed $\mathcal{N}\!=\!6$ Chern-Simons theory, are not allowed \cite{Ooguri:2009cv,Jeong:2009aa}. See also \cite{Donos:2009zf} for a study of enhanced supersymmetry in different class of solutions with Schr\"{o}dinger invariance. From our field theory study, it may perhaps be interesting to study spontaneous supersymmetry breaking from the gravity side \cite{Giveon:2009bv}.

The remaining part of this paper is organized as follows. In section 2 we review the mass-deformed $\mathcal{N}\!=\!6$ Chern-Simons-matter theory and explain its classical supersymmetric vacua. We explain them in the $\mathcal{N}\!=\!2$ formulation. We also add Yang-Mills term preserving this amount of supersymmetry, and also discuss the order and signs of masses for fluctuations around various vacua.
In section 3 we put this theory on a small 2-torus, explain the known result on Yang-Mills Chern-Simons index and finally derive our result. We also explain the map between our supersymmetric vacua and the fermion droplets of \cite{Lin:2004nb}.

\section{The theory and classical vacua}

The $\mathcal{N}\!=\!6$ Chern-Simons-matter theory consists of two $U(N)$ gauge fields $A_\mu$, $\tilde{A}_\mu$, four complex scalars $Z_a$, $\bar{W}^{\dot{a}}$ ($a,\dot{a}=1,2$) in bifundamental representation of $U(N)\times U(N)$, and four complex fermions $\psi_a,\bar\chi^{\dot{a}}$ which are superpartners of the scalars. See \cite{Benna:2008zy,Kim:2009wb} for our notation, which only shows manifest $\mathcal{N}\!=\!2$ supersymmetry. It was also shown in \cite{Hosomichi:2008jb} that one can introduce a mass deformation preserving $\mathcal{N}\!=\!6$ Poincare supersymmetry. The action is given by
\begin{eqnarray}
  \mathcal{L}&=&\mathcal{L}_{CS}+\mathcal{L}_M+\mathcal{L}_{sup}+\mathcal{L}_\mu\nonumber\\
  \mathcal{L}_{CS}&=&\frac{k}{4\pi}{\rm tr}\left[\left(AdA-\frac{2i}{3}A^2+i\bar\lambda\lambda
  -2D\sigma\right)
  -\left(\tilde{A}d\tilde{A}-\frac{2i}{3}\tilde{A}^2+i\bar{\tilde\lambda}\tilde\lambda
  -2\tilde{D}\tilde\sigma\right)\right]\nonumber\\
  \mathcal{L}_M&=&{\rm tr}\left[\frac{}{}\!\right.-D_\mu Z_aD^\mu\bar{Z}^a-D_\mu W_{\dot{a}}D^\mu\bar{W}^{\dot{a}}-i\bar\psi^a\gamma^\mu D_\mu\psi_a-i\bar\chi^{\dot{a}}
  \gamma^\mu D_\mu\chi_{\dot{a}}\nonumber\\
  &&\hspace{0.6cm}-\left(\sigma Z_a-Z_a\tilde\sigma\right)\left(\bar{Z}^a\sigma-\tilde\sigma\bar{Z}^a\right)
  -\left(\tilde\sigma W_{\dot{a}}-W_{\dot{a}}\sigma\right)\left(\bar{W}^{\dot{a}}\tilde\sigma-
  \sigma\bar{W}^{\dot{a}}\right)\nonumber\\
  &&\hspace{0.6cm}+\bar{Z}^aDZ_a-W_{\dot{a}}D\bar{W}^{\dot{a}}-Z_a\tilde{D}\bar{Z}^a
  +\bar{W}^{\dot{a}}\tilde{D}W_{\dot{a}}\nonumber\\
  &&\hspace{0.6cm}
  -i\bar\psi^a\sigma\psi_a+i\psi_a\tilde\sigma\bar\psi^a+i\bar{Z}^a\lambda\psi_a+i\bar\psi^a
  \bar\lambda Z_a-i\psi_a\tilde\lambda\bar{Z}^a-iZ_a\bar{\tilde\lambda}\bar\psi^a\nonumber\\
  &&\hspace{0.6cm}
  +i\chi_{\dot{a}}\sigma\bar\chi^{\dot{a}}-i\bar\chi^{\dot{a}}\tilde\sigma\chi_{\dot{a}}
  -i\chi_{\dot{a}}\lambda\bar{W}^{\dot{a}}-iW_{\dot{a}}\bar\lambda\bar\chi^{\dot{a}}
  +i\bar{W}^{\dot{a}}\tilde\lambda\chi_{\dot{a}}+i\bar\chi^{\dot{a}}\bar{\tilde\lambda}W_{\dot{a}}
  \left.\frac{}{}\!\right]\nonumber\\
  \mathcal{L}_{sup}&=&-\frac{2\pi}{k}\int d^2\theta\epsilon^{ab}\epsilon^{\dot{a}\dot{b}}
  {\rm tr}\left(Z_aW_{\dot{a}}Z_bW_{\dot{b}}\right)+c.c.\nonumber\\
  \mathcal{L}_\mu&=&-\frac{\mu}{2}{\rm tr}\left[D+\tilde{D}\right]\ .\label{lagrangian}
\end{eqnarray}
In the $\mathcal{N}\!=\!2$ formulation, the mass deformation can be written either as a Fayet-Iliopoulos term as shown above, or as an F-term deformation \cite{Gomis:2008vc}, which are shown to be equivalent after a field redefinition \cite{Kim:2009ny}. The adjoint scalars $\sigma,\tilde\sigma$ and fermions $\lambda,\tilde\lambda$ are auxiliary at this stage. In particular, the scalars are given by
\begin{equation}
  \frac{k}{2\pi}\sigma=Z_a\bar{Z}^a-\bar{W}^{\dot{a}}W_{\dot{a}}-\frac{\mu}{2}\ ,\ \
  \frac{k}{2\pi}\tilde\sigma=\bar{Z}^aZ_a-W_{\dot{a}}\bar{W}^{\dot{a}}+\frac{\mu}{2}
\end{equation}
from equations of motion of $D, \tilde{D}$.

The classical supersymmetric vacua of this theory are analyzed in \cite{Gomis:2008vc}. Assuming $\mu>0$, the classical vacua are given by a direct sum of the following irreducible rectangular blocks: the first type of blocks is
\begin{equation}\label{irreducible-1}
  Z_1=\mu^{\frac{1}{2}}\left(\begin{array}{cccccc}\sqrt{n\!-\!1}\!\!\!&0&&&&\\&\!\sqrt{n\!-\!2}
  \!\!&\!0&&&\\
  &&\ddots&\ddots&&\\&&&\sqrt{2}&0&\\&&&&1&0\end{array}\right),\
  Z_2=\mu^{\frac{1}{2}}\left(\begin{array}{cccccc}0&1&&&&\\&0&\sqrt{2}&&&\\
  &&\ddots&\ddots&&\\&&&0\!&\!\!\sqrt{n\!-\!2}\!&\\&&&&0&\!\!\!\sqrt{n\!-\!1}\end{array}\right)
\end{equation}
with $\bar{W}^{\dot{1}}\!=\bar{W}^{\dot{2}}\!={\bf 0}_{(n\!-\!1)\!\times\!n}$ ($n\!=\!1,2,\cdots$), and the second type is
\begin{equation}\label{irreducible-2}
  \bar{W}^{\dot{1}}=\mu^{\frac{1}{2}}\left(\begin{array}{ccccc}\sqrt{n}\!\!\!&&&&\\0&\!\sqrt{n\!-\!1}&&&\\
  &0&\ddots&&\\&&\ddots&\sqrt{2}&\\&&&0&1\\&&&&0\end{array}\right)\ ,\ \
  \bar{W}^{\dot{2}}=\mu^{\frac{1}{2}}\left(\begin{array}{ccccc}0&&&&\\1&0&&&\\
  &\sqrt{2}&\ddots&&\\&&\ddots&0&\\&&&\sqrt{n\!-\!1}\!&\!0\\&&&&\!\!\!\sqrt{n}\end{array}\right)
\end{equation}
with $Z_1\!=\!Z_2\!=\!{\bf 0}_{(n\!+\!1)\!\times\!n}$ ($n\!=\!1,2,\cdots$), up to global gauge transformations. $n$ in both types of blocks is the number of columns. The first block with $n\!=\!1$ is understood as an empty column. The nonzero scalars in the first block can be written as $Z_a=\mu^{\frac{1}{2}}\mathcal{M}_a^{(n)}$, while those in the second blocks can be written as $\bar{W}^{\dot{a}}=\mu^{\frac{1}{2}}(\mathcal{M}_a^{(n\!+\!1)})^\dag$. The `tracelss' parts (in $a,b$ indices) of $J_a^{\ b}=\mu^{-1}Z_a\bar{Z}^b$ and $\tilde{J}^a_{\ b}=\mu^{-1}\bar{Z}^aZ_b$ form representations of $SU(2)$, and similar for $W_{\dot{a}}$. The most general solution is a direct sum of these,
\begin{equation}\label{general-vacuum}
  Z_a=\mu^{\frac{1}{2}}\left(\begin{array}{c}
  \begin{array}{cccccc}\mathcal{M}_a^{(n_1)}\!\!&&&&&\\&\!\!\ddots\!&&&&\\
  &&\!\!\mathcal{M}_a^{(n_i)}&&&\\&&&0&&\\&&&&\ddots&\\&&&&&0\end{array}\\
  \left[\hspace{2.3cm}\begin{array}{c}{\bf 0}\\ \end{array}\hspace{2.3cm}\right]
  \end{array}\right)\ ,\ \
  \bar{W}^{\dot{a}}=\mu^{\frac{1}{2}}\left(\begin{array}{c}
  \begin{array}{cccccc}0&&&&&\\&\ddots&&&&\\&&0&&&\\
  &&&{\mathcal{M}_a^{(n_i\!+\!1)}}^\dag\!\!&&\\&&&&\!\!\ddots\!&\\
  &&&&&\!\!{\mathcal{M}_a^{(n_f)}}^\dag\end{array}\\
  \left[\hspace{2.3cm}\begin{array}{c}{\bf 0}\\ \end{array}\hspace{2.3cm}\right]
  \end{array}\right)
\end{equation}
where the rectangular matrices $[\bf{0}]$ at the bottom denote empty rows.
We emphasize that the irreducible blocks above are not square matrices. This is the only difference with \cite{Gomis:2008vc}, which added an empty row to our $\mathcal{M}_a^{(n)}$ to use square blocks only. Their general solution took the form of (\ref{general-vacuum}) without explicit empty rows as ours, but necessarily has an empty row or column per block. In our solution, the numbers of empty columns and rows are explicitly controlled by $\mathcal{M}_a^{(1)}$ in the first block and $[\bf{0}]$ in (\ref{general-vacuum}). The above solution is slightly more general in this sense, which will play a role in the vacuum counting.

\begin{figure}[t]
  \begin{center}
    \hspace*{1cm}
    \includegraphics[width=10cm]{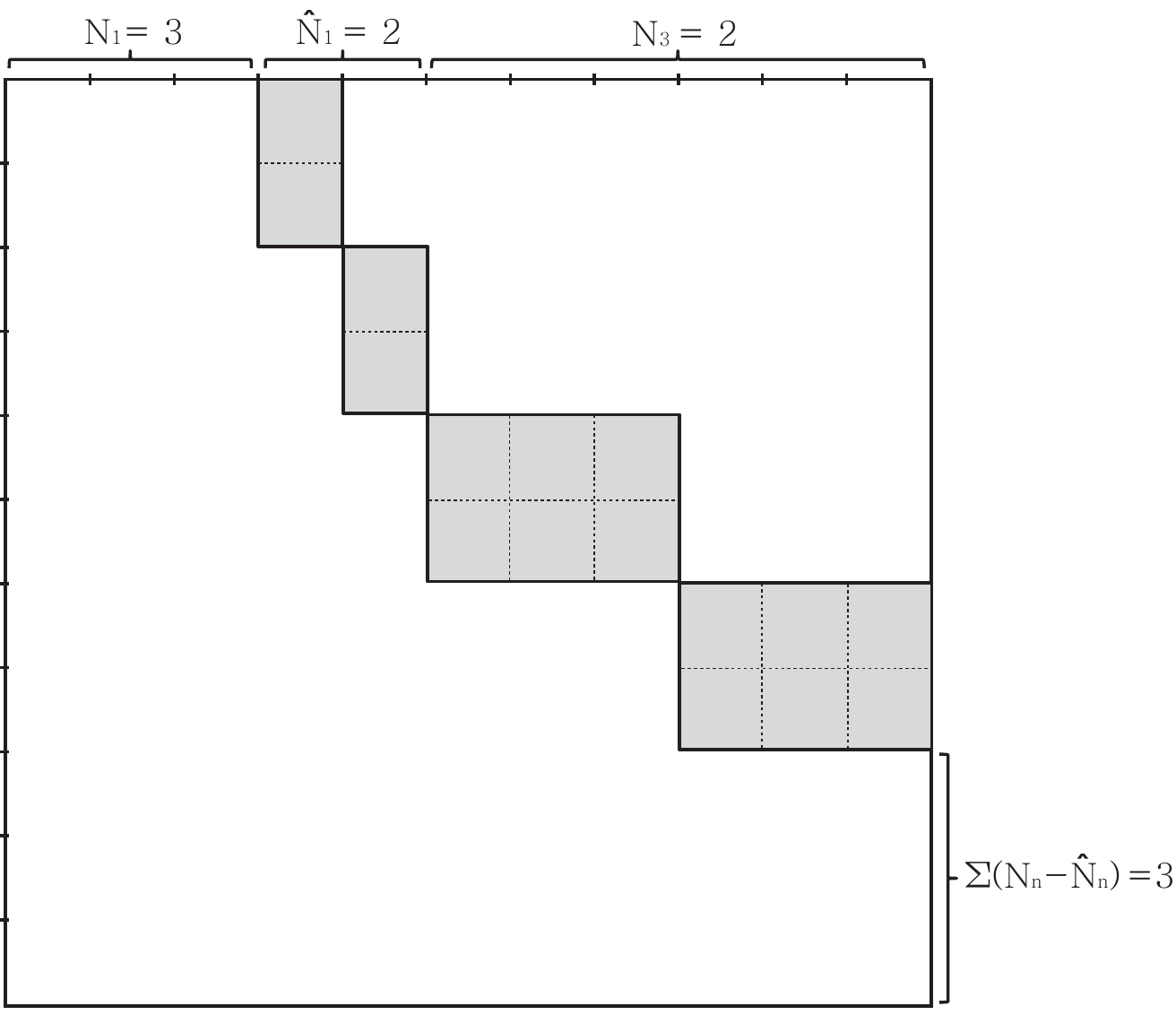}
      \caption{An example of our parametrization for $N\!=\!11$, with $N_1\!=\!3$, $\hat{N}_1\!=\!2$, $N_3\!=\!2$. Grey boxes with longer horizontal/vertical edges denote insertions of irreducible blocks of first/second type, respectively. Unbroken gauge group is $U(3)^2\times U(2)^2$ in this case.}\label{pic-vacuum}
  \end{center}
\end{figure}

We parametrize the above classical vacua as follows. Firstly, the summation of numbers of columns for all blocks is $N$, which is a partition of $N$. We further assign one of the two blocks to each integer in the partition. Let $N_n$ and $\hat{N}_n$ denote the number of first and second type of blocks with $n$ columns, respectively.
See figure \ref{pic-vacuum}.
These `occupation numbers' provide a labeled partition of $N$:
\begin{equation}\label{labeled-partition}
  \sum_{n=1}^\infty\left(nN_n+n\hat{N}_n\right)=N\ .
\end{equation}
In particular, the number of empty columns is given by $N_1$. On the other hand, the number of empty rows is given by $\sum_{n=1}^\infty\left(N_n-\hat{N}_n\right)$. Another constraint apart from (\ref{labeled-partition}) is
\begin{equation}\label{empth-row-1}
  0\leq\sum_{n=1}^\infty\left(N_n-\hat{N}_n\right)\leq N\ .
\end{equation}
It is easy to see that (\ref{labeled-partition}) always guarantees the second inequality in (\ref{empth-row-1}). So a more convenient set of constraints is (\ref{labeled-partition}) and
\begin{equation}\label{empty-row}
  \sum_{n=1}^\infty\left(N_n-\hat{N}_n\right)\geq 0\ .
\end{equation}
The scalars can be written as
\begin{equation}\label{vacuum-1}
  Z_a=\mu^{\frac{1}{2}}
  \bigoplus_{n=1}^\infty\left(\begin{array}{cc}[\mathcal{M}_a^{(n)}]_{(n\!-\!1)\times n}\otimes
  {\bf 1}_{N_n}&\\&{\bf 0}_{(n\!+\!1)\times n}\otimes{\bf 1}_{\hat{N}_n}\end{array}\right)
\end{equation}
and
\begin{equation}\label{vacuum-2}
  \bar{W}^{\dot{a}}=
  \mu^{\frac{1}{2}}\bigoplus_{n=1}^\infty\left(\begin{array}{cc}{\bf 0}_{(n\!-\!1)\times n}\otimes
  {\bf 1}_{N_n}&\\&[\mathcal{M}_a^{(n\!+\!1)}]^\dag_{(n\!+\!1)\times n}\otimes{\bf 1}_{\hat{N}_n}\end{array}\right)\ .
\end{equation}
The expectation values of the scalars $\sigma,\tilde\sigma$ are
\begin{equation}\label{vacuum-3}
  \frac{k}{2\pi}\sigma=\bigoplus_{n=1}^\infty\left(\begin{array}{cc}\mu n{\bf 1}_{(n\!-\!1)N_n}&\\
  &-\mu n{\bf 1}_{(n\!+\!1)\hat{N}_n}\end{array}\right)\oplus{\bf 0}_{\sum_{n}(N_n\!-\!\hat{N}_n)}
  -\frac{\mu}{2}{\bf 1}_N
\end{equation}
and
\begin{equation}\label{vacuum-4}
  \frac{k}{2\pi}\tilde\sigma=
  \bigoplus_{n=1}^\infty\left(\begin{array}{cc}\mu(n\!-\!1){\bf 1}_{nN_n}&\\
  &-\mu(n\!+\!1){\bf 1}_{n\hat{N}_n}\end{array}\right)+\frac{\mu}{2}{\bf 1}_N\ .
\end{equation}
The gauge symmetry unbroken by the above background is
\begin{equation}\label{unbroken}
  \bigotimes_{n=1}^\infty\left[U(N_n)\times U(\hat{N}_n)\right]\times
  U\left(\sum_n(N_n\!-\!\hat{N}_n)\right)\ ,
\end{equation}
from shuffling blocks with same sizes. The last factor comes from the empty rows.

For general $k$, the theory is strongly coupled and one should have a control over it to understand the quantum vacua. For the purpose of studying supersymmetric vacua, one can study the Witten index which is invariant under various deformations of the theory. We add the kinetic terms for the $U(N)\times U(N)$ vector supermultiplets
\begin{eqnarray}\label{yang-mills}
  \mathcal{L}_{YM}&=&\frac{1}{g^2}{\rm tr}\left[-\frac{1}{4}F_{\mu\nu}F^{\mu\nu}-\frac{1}{2}D_\mu\sigma
  D^\mu\sigma-i\bar\lambda\gamma^\mu D_\mu\lambda-i\bar\lambda[\sigma,\lambda]
  +\frac{1}{2}D^2\right.\nonumber\\
  &&\hspace{1.1cm}\left.-\frac{1}{4}\tilde{F}_{\mu\nu}\tilde{F}^{\mu\nu}-\frac{1}{2}D_\mu\tilde\sigma D^\mu\tilde\sigma-i\bar{\tilde\lambda}\gamma^\mu D_\mu\tilde\lambda
  -i\bar{\tilde\lambda}[\tilde\sigma,\tilde\lambda]+\frac{1}{2}\tilde{D}^2\right]
\end{eqnarray}
with a continuous constant $g^2$ having the dimension of mass. We will compute the index in the next section assuming $kg^2\ll\mu$. The classical solution (\ref{vacuum-1}), (\ref{vacuum-2}), (\ref{vacuum-3}), (\ref{vacuum-4}) is unaffected.

For later use, we consider the order of masses for various fluctuations around classical vacua, assuming the Yang-Mills deformation. The masses are affected by $g$ for $kg^2\ll\mu$. The final result is a natural one: all fields in the vector multiplets associated with the unbroken symmetry (\ref{unbroken}) acquire small masses of order $kg^2$ from the Chern-Simons terms, while the other fields acquire large masses which scale positively with $\mu$. More concretely, we will show that the masses for heavy modes scale either like $\mu/k$ or $\sqrt{\mu g^2}$.

We consider the fermion masses which are simpler. Bosonic masses follow from supersymmetry. The masses come from various fermion quadratures in (\ref{lagrangian}) and (\ref{yang-mills}).
To analyze these in components, we label the $N_n$ blocks of size $n$ by $i_n=1,2,\cdots,N_n$, and $\hat{N}_n$ blocks by $\hat{i}_n=1,2,\cdots,\hat{N}_n$. The rows and columns within each block of size $n$ are labeled by $\alpha_n$ or $\hat\alpha_n$, which run from $1$ to $n$, $n\!-\!1$ or $n\!+\!1$ as appropriate. The components connecting first type of blocks (i.e. nonzero $Z_a$) are written as
\begin{equation}\label{1-1-block}
  \left[\psi_a\right]_{\alpha_m i_m;\beta_n j_n}\ ,\ \ \left[\chi_{\dot{a}}\right]_{\alpha_m i_m;\beta_n j_n}
  \ ,\ \ \left[\lambda\right]_{\alpha_m i_m;\beta_n j_n}\ ,\ \ [\tilde\lambda]_{\alpha_mi_m;\beta_nj_n}\ ,
\end{equation}
where the first and last two indices are for rows and columns, respectively. There are also components connecting second type of blocks, and those connecting one first and one second type of blocks.  For the rows of $\psi_a$, $\bar\chi^{\dot{a}}$ and for the rows and columns of $\lambda$, we need additional index corresponding to empty rows in (\ref{general-vacuum}). Calling it $\alpha$, running from $1$ to $\sum_n(N_n\!-\!\hat{N}_n)$, we have components
\begin{equation}
  [\psi_a]_{\alpha;\beta_nj_n}\ ,\ \ [\bar\chi^{\dot{a}}]_{\alpha;\beta_nj_n}\ ,\ \
  [\lambda]_{\alpha;\beta}\ ,\ \ [\lambda]_{\alpha;\beta_nj_n}\ ,
\end{equation}
and also similar components replacing $\beta_nj_n$ by $\hat\beta_n\hat{j}_n$.

We first consider the components in (\ref{1-1-block}) connecting first type of blocks. After canonically normalizing the gauginos $\lambda,\tilde\lambda\rightarrow g\lambda, g\tilde\lambda$, one obtains
\begin{eqnarray}
  &&-i\frac{2\pi\mu(m\!-\!n)}{k}\left([\bar\lambda]_{\beta_n j_n;\alpha_m i_m}
  [\lambda]_{\alpha_m i_m;\beta_n j_n}+[\bar{\tilde\lambda}]_{\beta_n j_n;\alpha_m i_m}
  [\tilde\lambda]_{\alpha_m i_m;\beta_n j_n}-[\bar\psi^a]_{\beta_m j_m;\alpha_n i_n}
  [\psi_a]_{\alpha_n i_n;\beta_m j_m}\right)\nonumber\\
  &&+\frac{ikg^2}{4\pi}[\bar\lambda]_{\beta_n j_n;\alpha_m i_m}[\lambda]_{\alpha_m i_m;\beta_n j_n}
  -\frac{ikg^2}{4\pi}
  [\bar{\tilde\lambda}]_{\beta_n j_n;\alpha_m i_m}[\tilde\lambda]_{\alpha_m i_m;\beta_n j_n}\nonumber\\
  &&+i\sqrt{\mu g^2}[\mathcal{M}_a^{(m)}]^\dag_{\beta_m;\alpha_m}[\lambda]_{\alpha_m i_m;\gamma_n j_n}
  [\psi_a]_{\gamma_n j_n;\beta_m i_m}+
  i\sqrt{\mu g^2}[\bar\psi^a]_{\beta_mi_m;\gamma_n j_n}[\bar\lambda]_{\gamma_n j_n;\alpha_m i_m}[\mathcal{M}_a^{(m)}]_{\alpha_m\beta_m}\nonumber\\
  &&-i\sqrt{\mu g^2}[\psi_a]_{\alpha_n i_n;\gamma_m j_m}[\tilde\lambda]_{\gamma_m j_m;\beta_n i_n}
  [\mathcal{M}_a^{(n)}]^\dag_{\beta_n\alpha_n}-i\sqrt{\mu g^2}[\mathcal{M}_a^{(n)}]_{\alpha_n\beta_n}
  [\bar{\tilde\lambda}]_{\beta_n i_n;\gamma_m j_m}
  [\bar\psi^a]_{\gamma_m j_m;\alpha_n i_n}\label{1-1-mass1}
\end{eqnarray}
plus
\begin{eqnarray}\label{1-1-mass2}
  &&i\frac{2\pi\mu(m\!-\!n)}{k}[\bar\chi^{\dot{a}}]_{\beta_m j_m;\alpha_n i_n}
  [\chi_{\dot{a}}]_{\alpha_n i_n;\beta_m j_m}\\
  &&+\frac{2\pi\mu}{k}\left(\epsilon^{ab}\epsilon^{\dot{a}\dot{b}}
  [\mathcal{M}_a^{(m)}]_{\beta_m;\delta_m}[\chi_{\dot{a}}]_{\delta_mj_m;\gamma_ni_n}
  [\mathcal{M}_b^{(n)}]_{\gamma_n;\alpha_n}[\chi_{\dot{b}}]_{\alpha_n i_n;\beta_m j_m}
  +c.c.\right)\ .\nonumber
\end{eqnarray}
Summations over matrix indices as well as $m,n$ are understood. The first line of (\ref{1-1-mass1}) provides masses of order $\mu/k$, which is the biggest unless $m\!=\!n$. The third and fourth lines are of order $\sqrt{\mu g^2}$. The second line is of order $kg^2$.

We first study the order of masses for the coupled modes $\lambda, \tilde\lambda, \psi$ from (\ref{1-1-mass1}). When $m\!\neq\!n$, the leading masses for all fields are of order $\mu/k$ from the first line. When $m\!=\!n$, we have to study the zero modes from third and fourth line to see if they give dominant contribution to masses or if we have to consider the second line. The numbers of complex degrees of freedom in $\psi_a$, $\lambda$, $\tilde\lambda$ are
\begin{equation}
  2N_n^2(n^2\!-\!n)\ ,\ \ N_n^2(n\!-\!1)^2\ ,\ \ N_n^2n^2\ ,
\end{equation}
respectively. The sum of the latter two minus the former is $N_n^2$, which is exactly what we expect for light gauginos from enhanced $U(N_n)$ gauge symmetry. To exactly identify these modes, we consider the null vector condition for the gauginos
\begin{equation}\label{zero-mode}
  [\mathcal{M}_a^{(n)}]^\dag\lambda_{i_n;j_n}-[\tilde\lambda]_{i_n;j_n}[\mathcal{M}_a^{(n)}]^\dag=0
\end{equation}
where we suppressed the appearance of $\alpha_n$ type matrix indices. One can explicitly solve this equation and show that the general zero modes are given by
\begin{equation}\label{massless}
  \lambda={\bf 1}_{n\!-\!1}\otimes\Lambda_{N_n\times N_n}\ ,\ \ \tilde\lambda=
  {\bf 1}_n\otimes\Lambda_{N_n\times N_n}\ ,
\end{equation}
providing the expected $N_n^2$ complex modes. The mass of $\Lambda$ is $kg^2$ from the second line of (\ref{1-1-mass1}). The remaining $4N^2n(n\!-\!1)$ real nonzero modes have masses of order $\sqrt{\mu g^2}$. For the last nonzero modes, there are as many positive masses as there are negative ones, due to the off-diagonal structure of the last two lines between $\psi$ and $\lambda,\tilde\lambda$.

We also consider the $\chi$ masses from (\ref{1-1-mass2}). These masses are not affected at all by deforming with Yang-Mills like terms. So the fact that all fields have masses of order $\mu/k$ follows from the same fact in the original Chern-Simons-matter theory. We also checked it ourselves.

The mass analysis for the modes connecting the second type of blocks, like $[\psi_a]_{\hat\alpha_m\hat{i}_m;\hat\beta_n\hat{j}_n}$, is similar to the above, with the role of $\psi$, $\chi$ changed, as well as changing the role of $\lambda$, $\tilde\lambda$. In particular, we obtain the $\hat{N}_n^2$ light fermions for each unbroken $U(\hat{N}_n)$ symmetry, whose solution takes the form of (\ref{massless}) with $N_n$ replaced by $\hat{N}_n$, and ${\bf 1}_{n\!-\!1}$ replaced by ${\bf 1}_{n\!+\!1}$.

We then consider blocks connecting one first and one second type of blocks, like $[\psi_a]_{\alpha_mi_m;\hat\beta_n\hat{j}_n}$. There always come dominant mass terms from coupling to $\sigma,\tilde\sigma$ background, similar to the first lines of (\ref{1-1-mass1}) and (\ref{1-1-mass2}). From the structure of (\ref{vacuum-3}) and (\ref{vacuum-4}), these mass terms come with coefficient $\frac{2\pi\mu(m\!+\!n)}{k}$ instead of $\frac{2\pi\mu(m\!-\!n)}{k}$ in the previous cases, which is always nonzero.

We finally consider the blocks containing $\alpha$ type indices. For bi-fundamental matter modes $\psi_a$, $\bar\chi^{\dot{a}}$, the dominant masses of order $\mu/k$ again come from coupling to $\sigma,\tilde\sigma$. For gauginos $\lambda$, most of the modes acquire nonzero masses of order $\mu/k$ except $\lambda_{\alpha;\beta}$, which acquires mass of order $kg^2$ from Chern-Simons term. This is the expected light modes from unbroken $U\left(\sum_n(N_n\!-\!\hat{N}_n)\right)$ gauge symmetry in (\ref{unbroken}).

We close this section by summarizing the light mode contents and their effective action. From the light fermion modes of the form (\ref{massless}) corresponding to the unbroken symmetry (\ref{unbroken}), we also expect similar light modes from the adjoint bosons. In fact,
\begin{eqnarray}
  &&\hspace{-1.5cm}A_\mu={\bf 1}_{n\!-\!1}\otimes [a_{\mu}]_{N_n\times N_n},\
  \sigma={\bf 1}_{n\!-\!1}\otimes [\tau]_{N_n\times N_n};\
  \tilde{A}_{\mu}={\bf 1}_{n}\otimes[a_\mu]_{N_n\times N_n},\
  \tilde\sigma={\bf 1}_{n}\otimes [\tau]_{N_n\times N_n}\nonumber\\
  &&\hspace{-1.5cm}A_\mu={\bf 1}_{n\!+\!1}\otimes [a_{\mu}]_{\hat{N}_n\times\hat{N}_n},\
  \sigma={\bf 1}_{n\!+\!1}\otimes [\tau]_{\hat{N}_n\times\hat{N}_n};\
  \tilde{A}_{\mu}={\bf 1}_{n}\otimes[a_\mu]_{\hat{N}_n\times\hat{N}_n},\
  \tilde\sigma={\bf 1}_{n}\otimes [\tau]_{\hat{N}_n\times\hat{N}_n}\label{low-energy}\\
  &&\hspace{-1.5cm}A_\mu=[a_{\mu}]_{p\times p},\
  \sigma=[\tau]_{p\times p}\ \ \ \ ({\rm where}\ \ p\equiv\sum(N_n\!-\!\hat{N}_n)\ )\nonumber
\end{eqnarray}
corresponding to $U(N_n)$, $U(\hat{N}_n)$, $U\left(\sum(N_n-\hat{N}_n)\right)$, respectively, are the superpartner modes. $a_\mu, \tau, \Lambda$ form an $\mathcal{N}\!=\!2$ vector multiplet. One can integrate out the heavy modes whose masses scale either like $\mu/k$ or $\sqrt{\mu g^2}$, after which one obtains the $\mathcal{N}\!=\!2$ Yang-Mills Chern-Simons theory with gauge group $U(N_n)_{-k}$, $U(\hat{N}_n)_k$, $U\left(\sum(N_n\!-\!\hat{N}_n)\right)_k$ and the Yang-Mills coupling constant $\frac{g}{\sqrt{2n\!-\!1}}$, $\frac{g}{\sqrt{2n\!+\!1}}$, $g$, respectively. The subscripts for gauge groups denote the Chern-Simons levels. These values of Chern-Simons levels and Yang-Mills coupling can be obtained by inserting (\ref{low-energy}) to (\ref{lagrangian}) and (\ref{yang-mills}). For instance, considering the gauge fields on the first line of (\ref{low-energy}), one obtains
\begin{equation}
  \frac{k}{4\pi}\int\left({\rm tr}{\bf 1}_{n\!-\!1}\!-\!{\rm tr}{\bf 1}_n\right)
  {\rm tr}_{N_n}\left(ada-\frac{2i}{3}a^3\right)=-\frac{k}{4\pi}\int
  {\rm tr}_{N_n}\left(ada-\frac{2i}{3}a^3\right)\ .
\end{equation}
The integrated out heavy fermions do not shift the Chern-Simons levels for the unbroken gauge groups. To show this, it suffices to check that numbers of positive and negative masses are equal for all charged fermions under a $U(r)$ factor in (\ref{unbroken}). From the above analysis of masses, this structure is obvious for almost all modes. For instance, it is easy to see that $[\lambda]_{i_m;j_n}$ and $[\lambda]_{i_n;j_m}$ for $m\!\neq\!n$ always have masses of opposite signs, providing canceling contribution to the level shifts. Appropriate pairs of ($\psi_a$, $\bar\chi^{\dot{a}}$) have masses of opposite signs as well, except when the mass term involves superpotential, say as (\ref{1-1-mass2}). We briefly explain the structure for $\psi_a$ and $\chi_{\dot{a}}$ modes connecting first types of blocks (same argument applies to second type of blocks), which is the only nontrivial case. When $m\!=\!n$, $\psi_a$ modes gain masses of order $\sqrt{\mu g^2}$ by mixing with $\lambda,\tilde\lambda$, which was already shown to come with equal positive/negative masses. For $\chi_{\dot{a}}$, one can explicitly diagonalize the second line of (\ref{1-1-mass2}) by using $SU(2)$ Clebsch-Gordan analysis for the vacuum. The basic idea is to consider $\mathcal{M}\chi\mathcal{M}$ appearing in the mass term as a linear operator acting on $\chi$, whose eigenvalues can be obtained as square-roots of
$\mathcal{M}^\dag\mathcal{M}\chi\mathcal{M}\mathcal{M}^\dag$. This always provides pairs of positive/negative masses. For $m\!<\!n$, the modes $[\psi_a]_{i_m;j_n}$ and $[\psi_a]_{i_n;j_m}$ have opposite signs in their masses, but the numbers of complex degrees of freedom are $2N_mN_n(m\!-\!1)n$ and $2N_mN_n(n\!-\!1)m$, whose difference is
\begin{equation}\label{unbalance}
  2N_mN_n(n\!-\!m)\ .
\end{equation}
On the other hand, the $\chi_{\dot{a}}$ modes in (\ref{1-1-mass2}) can be diagonalized again using Clebsch-Gordan decomposition. Now there appears kernel of the operation
$\epsilon^{ab}\mathcal{M}_a^{(m)}[\chi_{\dot{a}}]_{i_m;j_n}\mathcal{M}^{(n)}_b$, whose complex dimension is exactly $2(n\!-\!m)$. The masses of these modes, of definite sign, come from the first line of (\ref{1-1-mass2}) and exactly compensates the unbalance of (\ref{unbalance}). The remaining masses involving nonzero-modes of this operator come in pairs of positive and negative masses.

\section{The index}

Following \cite{Witten:1999ds}, we consider the theory discussed in the previous section (with Yang-Mills kinetic term) compactified on a 2-torus with radii $r$ for two circles.
The Witten index ${\rm tr}(-1)^F$ that we would like to compute is independent of the continuous parameters $g, r, \mu$ in this theory.\footnote{This argument may be subtle in the limit $g,r\!\rightarrow\!\infty$ in general. We expect $g\rightarrow\infty$ to be safe since the mechanical model reduced on $T^2$ describes the motion of a `charged particle' with mass of order $g^{-2}$, subject to $k$ units of magnetic flux \cite{Witten:1999ds}. In the zero mass limit, we keep the lowest Landau level (supersymmetric states) while heavy states on the higher levels are irrelevant. On the other hand, the limit $r\!\rightarrow\!\infty$ is expected to be safe mostly for circumstantial reasons: the result on $T^2$ was used to successfully understand, among others, the s-rule of D-branes \cite{Ohta:1999iv} or the domain wall degeneracy \cite{Acharya:2001dz} in uncompactified cases.} The three mass scales $r^{-1},\mu,kg^2$ play important roles in our calculation. $r^{-1}$ is the energy of the Kaluza-Klein modes of fields along the 2-torus. $\mu$ is the mass appearing in our Lagrangian, setting the two mass scales $\mu/k$ and $\sqrt{\mu g^2}$. $kg^2$ is the bare mass of the gauge fields and their superpartners unbroken by a classical vacuum. Assuming generic $k$, and taking advantage of the fact that we can change the values of the mass scales without affecting the index,
we set
\begin{equation}
  r^{-1}\gg\frac{\mu}{k}\gg\sqrt{\mu g^2}\gg kg^2\ .
\end{equation}
We shall calculate the index by a `Born-Oppenheimer approximation.'
Firstly, we reduce the index calculation in the quantum field theory to that in the mechanics after compactification on a small torus \cite{Witten:1999ds}, ignoring all the Kaluza-Klein modes with masses of order $r^{-1}$.

After this reduction, we pick a classical vacuum, around which there appear modes with large masses scaling with $\mu$ and modes with small masses of order $kg^2$, as we explained in the previous section. We study the heavy modes first. The analysis of supersymmetric states for these modes is done by `Gaussian' (or quadratic) approximation, which reduces to the study of supersymmetric harmonic oscillators with frequencies given by their masses. The only issue is whether the supersymmetric ground state for a super-oscillator is bosonic or fermionic. It is well known that the ground state of the oscillator is bosonic/fermionic if the sign of the fermion mass is positive/negative. Let us denote by $M_P$ the total number of negative fermion masses for a given vacuum $P$. Then the contribution of heavy modes to the index is $(-1)^{M_P}$. $M_P$ is basically the Morse index associated with a critical point $P$ of the Morse function in supersymmetric quantum mechanics, studied by Witten \cite{Witten:1982im}. The essential point in \cite{Witten:1982im} which enabled the above quadratic approximation to provide the exact index was the existence of a nilpotent operator, providing a cohomology structure to the Hilbert space. In our $\mathcal{N}\!=\!2$ formulation, we keep four supersymmetries $Q_\alpha$, $\bar{Q}_{\alpha}$ manifest. Any of them is nilpotent off-shell, so we expect this argument to be true in our case.

One can easily see that $M_P$ is always even in our case. Firstly, recall that we showed at the end of section 2 that the heavy modes whose masses scale with $\mu$ come in pairs of positive and negative masses. So it suffices to show that the total number of heavy Majorana fermions is a multiple of $4$ (i.e. even complex fermions). The total numbers of complex fermions from matters $\psi_a,\chi_{\dot{a}}$ and gauginos $\lambda,\tilde\lambda$ are $4N^2$ and $2N^2$, respectively, so it suffices to show that the number of complex light fermions is even. The last number is given by
\begin{equation}
  \sum_{n=1}^\infty\left(N_n^2+\hat{N}_n^2\right)+\left(\sum_{n=1}^\infty(N_n\!-\!\hat{N}_n)\right)^2
  =2\sum_{n=1}^\infty\left(N_n^2+\hat{N}_n^2\right)+2\sum_{m<n}\left(N_mN_n+\hat{N}_m\hat{N}_n\right)
  -2\sum_{m,n=1}^\infty N_m\hat{N}_n\ ,
\end{equation}
from the unbroken gauge symmetry (\ref{unbroken}). This is indeed even, which proves $(-1)^{M_P}=1$.

Finally, we should consider the dynamics of modes with masses of order $kg^2$. As explained in the last paragraph of the previous section, the mechanical system that we have is the $T^2$ reduction of $\mathcal{N}\!=\!2$ Yang-Mills Chern-Simons theory with gauge group and levels given by
\begin{equation}\label{unbroken-again}
  \bigotimes_{n=1}^\infty\left[U(N_n)_{-k}\times U(\hat{N}_{n})_k\right]
  \times U\left(\sum(N_n\!-\!\hat{N}_n)\right)_k\ .
\end{equation}
The degrees of freedom associated with different factors of unitary gauge group do not interact, so that we can simply multiply the index computed in each part. Indices for $\mathcal{N}\!=\!2$ Yang-Mills Chern-Simons theories at level $\pm k$ with $SU(r)$ and $U(r)$ gauge groups are studied in \cite{Ohta:1999iv} and \cite{Acharya:2001dz}, respectively. The index for $U(r)$ is
\begin{equation}\label{YM-CS index}
  \left(\begin{array}{c}|k|\\r\end{array}\right)\equiv\frac{|k|!}{r!(|k|\!-\!r)!}
\end{equation}
for $r\leq|k|$, and zero otherwise.\footnote{Overall sign is always positive here, in contrast to the $\mathcal{N}\!=\!1$ case where $(-1)^r$ appears if $k\!<\!0$ \cite{Witten:1999ds}.} It is believed that the vanishing of the index for $r\!>\!|k|$ implies dynamical supersymmetry breaking. Combining the results for different gauge groups, one obtains
\begin{equation}\label{degeneracy}
 \prod_{n=1}^\infty\left[\left(\begin{array}{c}|k|\\N_n\end{array}\right)
  \left(\begin{array}{c}|k|\\\hat{N}_n\end{array}\right)\right]
  \left(\begin{array}{c}|k|\\ \sum(N_n\!-\!\hat{N}_n)\end{array}\right)
\end{equation}
for each classical vacuum parametrized by $N_n,\hat{N}_n$, satisfying (\ref{labeled-partition}) and (\ref{empty-row}).

Let us compute the generating function
\begin{equation}
  I_k(q)\equiv\sum_{N=1}^\infty I_{kN}q^N
\end{equation}
for the index $I_{kN}$, introducing a chemical potential $q$. To conveniently keep track of the condition  (\ref{empty-row}) later, we also introduce another chemical potential $z$, and define a quantity
\begin{equation}
  I_k(q,z)\equiv{\rm tr}\left[(-1)^F q^{\sum_n(nN_n+n\tilde{N}_n)}z^{\sum_n(N_n-\tilde{N}_n)}
  \right]\ ,
\end{equation}
where the trace denotes summation over the nonnegative occupation numbers $N_n$ and $\tilde{N}_n$ without imposing (\ref{labeled-partition}), but subject to (\ref{empty-row}), taking into account the degeneracy (\ref{degeneracy}). The actual generating function $I_k(q)$ is obtained from this by
\begin{equation}
  I_k(q)=\int_0^{2\pi}\frac{d\theta}{2\pi}\left(\!\!\frac{}{}\right.\sum_{p=0}^{|k|}e^{-ip\theta}
  \left.\frac{}{}\!\!\right)I_k(q,e^{i\theta})\ .
\end{equation}
The summation in the parenthesis stops at $p\!=\!|k|$, since the index for $U(p)_k$ Yang-Mills Chern-Simons theory is zero beyond this value. More concretely, one finds
\begin{equation}
  I_k(q)=\sum_{p=0}^{|k|}\int_0^{2\pi}\frac{d\theta}{2\pi}\ e^{-ip\theta}
  \sum_{N_n,\hat{N}_n=0}^{|k|}\left(\begin{array}{c}|k|\\ \sum(N_n\!-\!\hat{N}_n)
  \end{array}\right)\prod_{n=1}^\infty\left[\left(\begin{array}{c}|k|\\N_n\end{array}\right)
  \left(\begin{array}{c}|k|\\\hat{N}_n\end{array}\right)q^{nN_n\!+\!n\hat{N}_n}\right]
  e^{i\theta\sum(N_n\!-\!\hat{N}_n)}\ .
\end{equation}
All summations of $N_n,\hat{N}_n$ stop at $|k|$ again from dynamical supersymmetry breaking.
One might think that the summation over $N_n,\hat{N}_n$ should be restricted to those satisfying (\ref{empty-row}). However, this is already taken care of by the $\theta$ integration, which yields
\begin{equation}\label{delta}
  \int_0^{2\pi}\frac{d\theta}{2\pi}e^{i\theta(\sum(N_n\!-\!\hat{N}_n)-p)}=
  \delta_{\sum(N_n\!-\!\hat{N}_n)\ ,\ p}
\end{equation}
with positive $p$. From (\ref{delta}), one can replace $\sum(N_n\!-\!\hat{N}_n)$ appearing in the combinatoric factor by $p$ and obtain
\begin{equation}
  I_k(q)=\sum_{p=0}^{|k|}\int_0^{2\pi}\frac{d\theta}{2\pi}\ e^{-ip\theta}
  \left(\begin{array}{c}|k|\\p\end{array}\right)\sum_{N_n,\hat{N}_n=0}^{|k|}
  \prod_{n=1}^\infty\left[\left(\begin{array}{c}|k|\\N_n\end{array}\right)\left(\begin{array}{c}
  |k|\\\hat{N}_n\end{array}\right)(q^ne^{i\theta})^{N_n}(q^ne^{-i\theta})^{\hat{N}_n}\right]\ ,
\end{equation}
where the summation over $N_n,\hat{N}_n$ is still unrestricted due to (\ref{delta}).
Now one finds that the summations over $N_n$ and $\hat{N}_n$ all factorize: after using the binomial expansion formula, one obtains the following simple expression
\begin{equation}
  I_k(q)=\int_0^{2\pi}\frac{d\theta}{2\pi}\left[\left(1+e^{-i\theta}\right)
  \prod_{n=1}^\infty\left(1+q^ne^{i\theta}\right)\left(1+q^ne^{-i\theta}\right)\right]^{|k|}\ .
\end{equation}
The quantity appearing in $\left[\ \ \right]^{|k|}$ can be simplified using the Jacobi's triple product identity, in the form which is useful in studying bosonization of 2 dimensional QFT \cite{Green:1987sp}:
\begin{equation}
  2q^{\frac{1}{8}}\cos\frac{\theta}{2}\prod_{n=1}^\infty
  (1-q^n)(1+q^ne^{i\theta})(1+q^ne^{-i\theta})=\sum_{n=-\infty}^\infty q^{\frac{1}{2}\left(n\!+\!\frac{1}{2}\right)^2}e^{i\left(n\!+\!\frac{1}{2}\right)\theta}\ .
\end{equation}
Using this, one obtains the index
\begin{equation}\label{jacobi}
  I_k(q)=\int_0^{2\pi}\frac{d\theta}{2\pi}\left[\prod_{n=1}^\infty
  \frac{1}{1-q^n}\sum_{n=-\infty}^\infty
  q^{\frac{n(n\!+\!1)}{2}}e^{in\theta}\right]^{|k|}=\prod_{n=1}^\infty\frac{1}{(1-q^n)^{|k|}}\ \sum_{\substack{n_1,n_2,\cdots,n_{|k|}=-\infty\\(n_1+n_2+\cdots+n_{|k|}=0)}}^\infty q^{\frac{1}{2}\left(n_1^2+n_2^2+\cdots+n_{|k|}^2\right)}\ ,
\end{equation}
which is the asserted expression (\ref{general-partition}).

As explained in the introduction, this expression immediately reproduces the partitions of $N$ for $k\!=\!1$, perfectly agreeing with the expectation from the gravity dual. The underlying structure can be understood formally from bosonization. For $k\!=\!1$, the unbroken gauge symmetry of supersymmetric vacua should be no larger than $U(1)$, since $U(r)$ $\mathcal{N}\!=\!2$ Yang-Mills Chern-Simons theory at level $k$ admits supersymmetric vacua only for $r\leq|k|$. So all the occupation numbers $N_n$ and $\hat{N}_n$ effectively behave like those for `fermions,' taking $0$ or $1$.\footnote{Of course we do not mean fermions in the statistical sense.} Now the partition function for the supersymmetric vacua is like the contribution from two chiral fermions in 2 dimension. The Jacobi's identity that we used is nothing but an equality of 1-loop partition function of 2 chiral fermions and that of a chiral boson \cite{Green:1987sp}.

One can see more explicitly that the above fermions are nothing but the fermions appearing in the half BPS gravity solutions of \cite{Lin:2004nb}. To clearly see this, let us rename the $\hat{N}_n$ occupation numbers as follows: define $N^\prime_n=\hat{N}_{n\!-\!1}$ for $n\geq 2$, i.e. $n$ in  $N^\prime_n$ denotes the number of rows of the block of second type. $N^\prime_n$ for $n\geq 2$ are what we considered before. We also introduce $N^\prime_1$, which is now the number of empty rows. From the discussions before (\ref{empty-row}), we find that $N^\prime_1=\sum_{n=1}(N_n\!-\!\hat{N}_n)$. We can parametrize our supersymmetric vacua by two sets of fermionic occupation numbers $N_n,N^\prime_n$ (with $n\geq 1$) satisfying
\begin{equation}\label{antiparticle}
  \sum_{n=1}^\infty N_n=\sum_{n=1}^\infty N^\prime_n\ .
\end{equation}
This is nothing but the condition in \cite{Lin:2004nb} that the `$U(1)$ charge' is set to zero, namely the Fermi level of the droplet should be such that we have equal number of particles and holes. Having $N_n\!=\!1$ corresponds to exciting a droplet of unit area in \cite{Lin:2004nb} whose height is $n\!-\!1$ from the Fermi level, while $N^\prime_n\!=\!1$ corresponds to having an empty area (white region) with depth $n\!-\!1$ beneath the Fermi level. In this fashion, we can concretely map each supersymmetric vacuum to the corresponding gravity solution. In particular, we can say which vacua have gravity duals with small curvatures (in which case we can do reliable gravity calculations). It should be interesting to consider such vacua and study them holographically, like \cite{Auzzi:2009es}, for instance.
  
For $k\!=\!2$, the infinite sum in (\ref{jacobi}) is given by
\begin{equation}\label{jacobi-2nd}
  \sum_{n=-\infty}^\infty q^{n^2}=\prod_{n=1}^\infty(1-q^{2n})(1+q^{2n\!-\!1})^2\ ,
\end{equation}
which yields the expression for $I_2(q)$ in (\ref{higher-partition}). For $k\!=\!3$, one enouncters
\begin{equation}\label{sum-k3}
  \sum_{m,n=-\infty}^\infty q^{m^2+n^2+mn}=\sum_{m=-\infty}^\infty
  q^{\frac{3}{4}m^2}\sum_{n=-\infty}^\infty q^{\left(n\!+\!\frac{m}{2}\right)^2}\ .
\end{equation}
We first consider the summation over $n$, with $m$ fixed. For even $m$, one uses (\ref{jacobi-2nd}): further summation over even $m$ can be changed into infinite product using the same formula, replacing $q$ by $q^3$. For odd $m$, the summation over $n$ is
\begin{equation}\label{jacobi-3rd}
  \sum_n q^{\left(n\!+\!\frac{1}{2}\right)^2}=
  2q^{\frac{1}{4}}\prod_{n=1}^\infty(1-q^{2n})(1+q^{2n})^2
\end{equation}
which is obtained by inserting $\theta\!=\!0$ in (\ref{jacobi}). Collecting all, one obtains
the expression for $I_3(q)$ in (\ref{higher-partition}). It is not clear to us which expression will be more physically suggestive between the infinite sum in (\ref{sum-k3}) or the product form (\ref{higher-partition}), but the product form makes it clear that one factor of (\ref{partition}) appears for $k\!\geq\!2$. Similar rearrangement can be made for larger values of $k$.

We also comment on the totally symmetric vacuum, where all scalars $Z_1,Z_2,W_{\dot{1}},W_{\dot{2}}$ are zero. After integrating out the massive matter fields, we have to consider an $\mathcal{N}\!=\!2$ Yang-Mills Chern-Simons theory with gauge group $U(N)\times U(N)$ at levels $k$ and $-k$. Since the index for this theory is zero for $N>|k|$, it is likely that the supersymmetry of this vacuum is spontaneously broken. This implies that the classical analysis of nonrelativistic superconformal Chern-Simons theory \cite{Nakayama:2009cz} based on this vacuum should acquire serious non-perturbative correction for $N>k$, at least as long as the symmetry of the vacuum is concerned. This may be a field theory explanation of the observation in \cite{Ooguri:2009cv,Jeong:2009aa} that gravity solutions with Schr\"{o}dinger invariance and $\mathcal{N}\!=\!6$ supersymmetry could not be found.

\vskip 0.7cm

\hspace*{-0.7cm}{\large\bf Acknowledgements}

\vskip 0.2cm

\hspace*{-0.75cm} We are grateful to Jerome Gauntlett, Ki-Myeong Lee, Sangmin Lee, Sungjay Lee, Jaemo Park, Soo-Jong Rey, Takao Suyama and Piljin Yi for helpful discussions. S.K. is supported in part by the Research Settlement Fund for the new faculty of Seoul National University. H.-C.K. is supported in part by the National Research Foundation of Korea (NRF) Grants No. 2007-331-C00073, 2009-0072755 and 2009-0084601.

\end{document}